\documentclass{PoS}

\usepackage{aas_macros}

\newcommand{\hess}{H.E.S.S.}
\newcommand{\rx}{RX\,J1713--3946}

\newcommand{\naima}{\textsf{naima}}

\usepackage{txfonts}
\usepackage[scaled=0.9]{helvet}

\title{\naima: a Python package for inference of relativistic particle energy
    distributions from observed nonthermal spectra}

\ShortTitle{\naima: inference of relativistic particle energy distributions}

\author{\speaker{V.~Zabalza}\\
        Department of Physics and Astronomy, University of
        Leicester, University Road, Leicester, LE1 7RH, United Kingdom\\
        E-mail:  \email{victor.zabalza@le.ac.uk}}

\abstract{ The ultimate goal of the observation of nonthermal emission from
    astrophysical sources is to understand the underlying particle acceleration
    and evolution processes, and few tools are publicly available to infer the
    particle distribution properties from the observed photon spectra from X-ray
    to VHE gamma rays. Here I present \naima, an open source Python package that
    provides models for nonthermal radiative emission from homogeneous
    distribution of relativistic electrons and protons. Contributions from
    synchrotron, inverse Compton, nonthermal bremsstrahlung, and neutral-pion
    decay can be computed for a series of functional shapes of the particle
    energy distributions, with the possibility of using user-defined particle
    distribution functions. In addition, \naima\ provides a set of functions that
    allow to use these models to fit observed nonthermal spectra through an MCMC
    procedure, obtaining probability distribution functions for the particle
    distribution parameters. Here I present the models and methods available in
    \naima\ and an example of their application to the understanding of a
    galactic nonthermal source.

\naima's documentation, including how to install the package, is available at
\texttt{http://naima.readthedocs.org}.}

\FullConference{The 34th International Cosmic Ray Conference,\\
		30 July- 6 August, 2015\\
		The Hague, The Netherlands}

\begin{document}
\section{Introduction}

Over the past few years, there has been a wealth of facilities that allow for an
unprecedented level of sensitivity in the high-energy and very-high-energy
gamma-ray bands, most notably \emph{Fermi}-LAT and the Cherenkov telescope
arrays \hess, MAGIC, and VERITAS. These energy bands probe the highest energies
at which particles are known to be accelerated, and provide valuable insight
into the acceleration and energy loss processes of a growing population of
galactic and extragalactic sources.

A first step in understanding the mechanisms through which a given source has
accelerated the particles that are responsible for the emission we detect is to
characterize the present age particle energy distribution. For many sources,
such as pulsar wind nebulae emitting gamma-rays through inverse Compton
scattering on the Cosmic Microwave Background (CMB), the derivation of the
particle distribution can be done in a model-independent way. To date, there are
few public codes that allow this sort of analysis, and in many papers the
discussion of observed spectra is carried out based on modelling code that is
not published along with the paper, therefore hampering the reproducibility of
the research.

Here I present \naima, an open source Python package developed with the aim of
providing proven methods for computing the nonthermal radiative output of
relativistic particle distributions. Being open source makes it easy for anyone
to check the algorithms used in the calculations, and of improving and extending
the package. Python is increasingly gaining ground as the language of choice for
astronomical research, and makes it easy to combine existing libraries such as
\naima\ into a given analysis.

There are two main components of the package: a set of nonthermal radiative
models, and a set of utility functions that make it easier to fit a given model
to observed spectral data.

\section{Features}

The radiative models are implemented in a modular way that allows to select a
functional shape for the particle distribution. There are several functions
included for this purpose, and the user can as well define their own following
the instructions in the documentation, therefore allowing to implement any type
of particle cooling, escape, or acceleration physics before computing its
radiative output with \naima. The radiative models currently available in \naima
are synchrotron, inverse Compton, nonthermal bremsstrahlung, and neutral pion
decay.  See Section \ref{sec:radiative} for details on the implementation. In
addition, \naima\ includes a set of wrappers around these models that allow them
to be used within the \textsf{sherpa} spectral analysis package
\cite{2001SPIE.4477...76F} in the module \texttt{naima.sherpa\_models}.

A set of fitting utilities are provided in \naima\ with the goal of infering the
properties of the parent particle distribution that gives rise to an observed
nonthermal spectrum. These use Markov Chain Monte Carlo (MCMC) sampling to obtain both
the maximum likelihood parameters as well as their uncertainties in a single
run. Details on the implementation can be found in Section~\ref{sec:mcmc}.
Several plotting functions are available as well to analyse the results of the
MCMC run. 

\naima\ uses the \texttt{astropy} package \cite{2013A&A...558A..33A} extensively,
most notably through the use of its physical unit module \texttt{astropy.units}.
This allows users to define the input spectra and parameters in their preferred
units, and \naima\ will be able to convert them as needed in its internal
calculations, with the added benefit of ensuring that the algorithms in \naima
are dimensionally correct.

\section{Implementation of radiation models} \label{sec:radiative}

\subsection{Synchrotron}
Synchrotron radiation is produced by all charged particles in the presence of
magnetic fields, and is ubiquitous in the emitted spectrum of leptonic sources.
A full description and derivation of its properties can be found in
\cite{1970RvMP...42..237B}.

The \texttt{naima.models.Synchrotron} class
implements the pa\-ra\-me\-tri\-za\-tion of the emissivity function of synchrotron
radiation in random magnetic fields presented by \cite[appendix
D]{2010PhRvD..82d3002A}. This parametrization is particularly useful as it
avoids using special functions, and achieves an accuracy of 0.2\
entire range of emission energy.

\subsection{Inverse Compton}

The inverse Compton (IC) scattering of soft photons by relativistic electrons is
the main gamma-ray production channel for electron populations
\cite{1970RvMP...42..237B}. Often, the seed photon field will be a blackbody or a diluted
blackbody, and the calculation of IC must be done taking this into account.
\naima\ implements the analytical approximations to IC upscattering of
blackbody radiation developed by \cite{2014ApJ...783..100K}. These have the
advantage of being computationally cheap compared to a numerical integration
over the spectrum of the blackbody, and remain accurate within one percent over
a wide range of energies. Both the isotropic IC and anisotropic IC
approximations are available in \naima. Note that even environments with complex
broadband background radiation can be effectively modelled by a combination of
diluted blackbodies.

The implementation in \naima\ allows to specify which blackbody seed photon
fields to use in the calculation, and provides the three dominant galactic
photon fields at the location of the Solar System through the CMB (Cosmic
Microwave Background), FIR (far-infrared dust emission), and NIR
(near-infrared stellar emission) keywords.  

\subsection{Nonthermal Bremsstrahlung}

Nonthermal bremsstrahlung radiation arises when a population of relativistic
particles interact with a thermal particle population. For the computation of the bremsstrahlung emission spectrum, the
\texttt{Bremsstrahlung} class implements the approximation of
\cite{1999ApJ...513..311B} to the exact cross-section presented by
\cite{1975ZNatA..30.1099H}.
Electron-electron bremsstrahlung is implemented for the complete energy range,
whereas electron-ion bremsstrahlung is at the moment only available for photon
energies above 10 MeV. The normalization of the emission, and importance of the
electron-electron versus the electron-ion channels, can be selected in the
class, with the default values assuming a fully ionised target medium with solar
abundances.
\subsection{Pion Decay}

The main gamma-ray production for relativistic protons are p-p interactions
followed by pion decay, which results in a photon with $E_\gamma >
100\,\mathrm{MeV}$. Until recently, the only parametrizations available for the
integral cross-section and photon emission spectra were either only applicable
to limited energy ranges \cite[$E_\gamma>0.1$\,TeV]{2006PhRvD..74c4018K}, or
were given as extensive numerical tables \cite{2006ApJ...647..692K}.  By
considering Monte Carlo results and a compilation of accelerator data on p-p
interactions, \cite{2014PhRvD..90l3014K} were able to extend the parametrization
of \cite{2006PhRvD..74c4018K} down to the energy threshold for pion production.
The \texttt{PionDecay} class uses an implementation of the formulae presented in
their paper, and gives the choice of which high-energy model to use (from the
parametrization to the different Monte Carlo results) through the
\texttt{hiEmodel} parameter. 

\section{MCMC sampling} \label{sec:mcmc}

The following will briefly describe the implementation of spectral fitting in
\naima, and a full explanation of MCMC and the sampling algorithm can be found
in \cite{2003itil.book.....M}, and in the documentation of \textsf{emcee}, the
package used for MCMC sampling \cite{2013PASP..125..306F}.

The measurements and uncertainties in the observed spectrum are assumed to be
correct, Gaussian, and independent (note that this is unlikely to be the case,
see Section~\ref{sec:future} on how this might be tackled in
the future).  Under this assumption, the likelihood of observed data given the
spectral model $S(\vec{p};E)$, for a parameter vector $\vec{p}$, is

\begin{equation}
    \mathcal{L} = \prod^N_{i=1} \frac{1}{\sqrt{2 \pi \sigma^2_i}} 
                \exp\left(-\frac{(S(\vec{p};E_i) - F_i)^2}{2\sigma^2_i}\right),
\end{equation}

where $(F_i,\sigma_i)$ are the flux measurement and uncertainty at an
energy $E_i$ over $N$ spectral measurements. Taking the logarithm,

\begin{equation}
    \ln\mathcal{L} = K - \sum^N_{i=1} \frac{(S(\vec{p};E_i) - F_i)^2}{2\sigma^2_i}.
\end{equation}

Given that the MCMC procedure will sample the areas of the distribution with
maximum value of the objective function, it is useful to define the objective
function as the log-likelihood disregarding constant factors:

\begin{equation}
    \ln\mathcal{L} \propto  \sum^N_{i=1} \frac{(S(\vec{p};E_i) - F_i)^2}{\sigma^2_i}.
\end{equation}

The $\ln\mathcal{L}$ function in this assumption can be related to the
$\chi^2$ parameter as $\chi^2=-2\ln\mathcal{L}$, so that
maximization of the log-likelihood is equivalent to a minimization of
$\chi^2$.

In addition to the likelihood from the observed spectral points, a prior
likelihood factor should be considered for all parameters. This prior likelihood
encodes our prior knowledge of the probability distribution of a given model
parameter. The combination of the prior and data likelihood functions is passed
onto the \textsf{emcee} sampler function, and the MCMC run is started.
\textsf{emcee} uses an affine-invariant MCMC sampler that has the advantage of
being able to sample complex parameter spaces without any tuning required. In
addition, having multiple simultaneous \emph{walkers} improves the efficiency of
the sampling and reduces the number of computationally-expensive likelihood
calls required.

\section{Example analysis: hadronic emission from \rx}\label{sec:rx}

As an example of nonthermal spectral analysis with \naima, here we will show the
results of inferring the particle distribution parameters of a hadronic
population from a spectrum of the shell-type supernova remnant \rx\ obtained
with the \hess\ Cherenkov telescope array \cite{2007A&A...464..235A}. The photon
spectrum in the 0.3--100\,TeV energy range is well characterized by a power-law
with an exponential cutoff, so we will use a similar function for the particle
distribution. Even though the leptonic or hadronic origin of the gamma-ray
emission from \rx\ is still under debate \cite[and references
therein]{2014MNRAS.445L..70G}, for the sake of example here we will assume an
neutral pion decay origin of the VHE gamma-ray emission. Several other examples,
with the full source code needed to reproduce them, are available in the \naima
documentation \texttt{http://naima.readthedocs.org}.

\begin{figure}[tb]
    \begin{center}
        \includegraphics[width=0.9\textwidth]{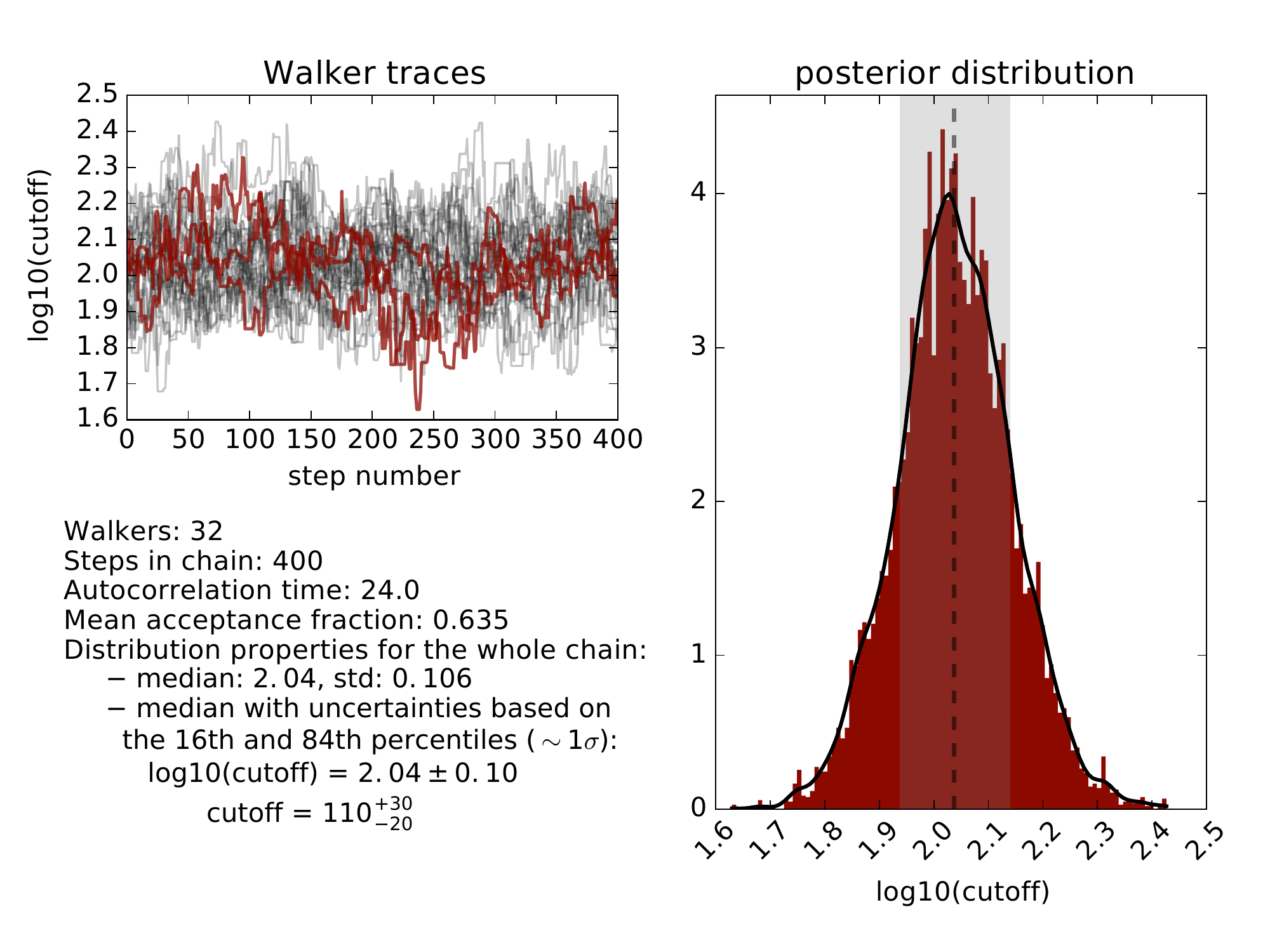}
    \end{center}
    \vspace{-2em}
    \caption{
        Diagnostic plot produced by \texttt{naima.plot\_chain} for the sampling
        of the cutoff energy in the particle distribution. The top-left panel
        shows the traces for the 32 walkers in gray, with three of them
        highlighted in red, that can be used to estimate whether the sampling
        has stabilized around the maximum likelihood parameters. The right panel
        shows the posterior distribution of the parameter, along with the median
        (dashed line) and an estimate of the 1$\sigma$ confidence interval (gray
        band). On the bottom-left, statistics of the parameter distribution,
        including a median with uncertainties based on the 16th and 84th
        percentiles, is shown.  Note that the parameter is sampled in
        logarithmic space in this example, and given the label
        \texttt{log10(cutoff)} \naima\ can identify this and convert it to its
        value and uncertainty in TeV, resulting in the estimate
        $E_\mathrm{cutoff} = 110^{+30}_{-20}$\,TeV.
    }
    \label{fig:chain}
\end{figure}

\begin{figure}[tb]
    \begin{center}
        \includegraphics[width=0.6\textwidth]{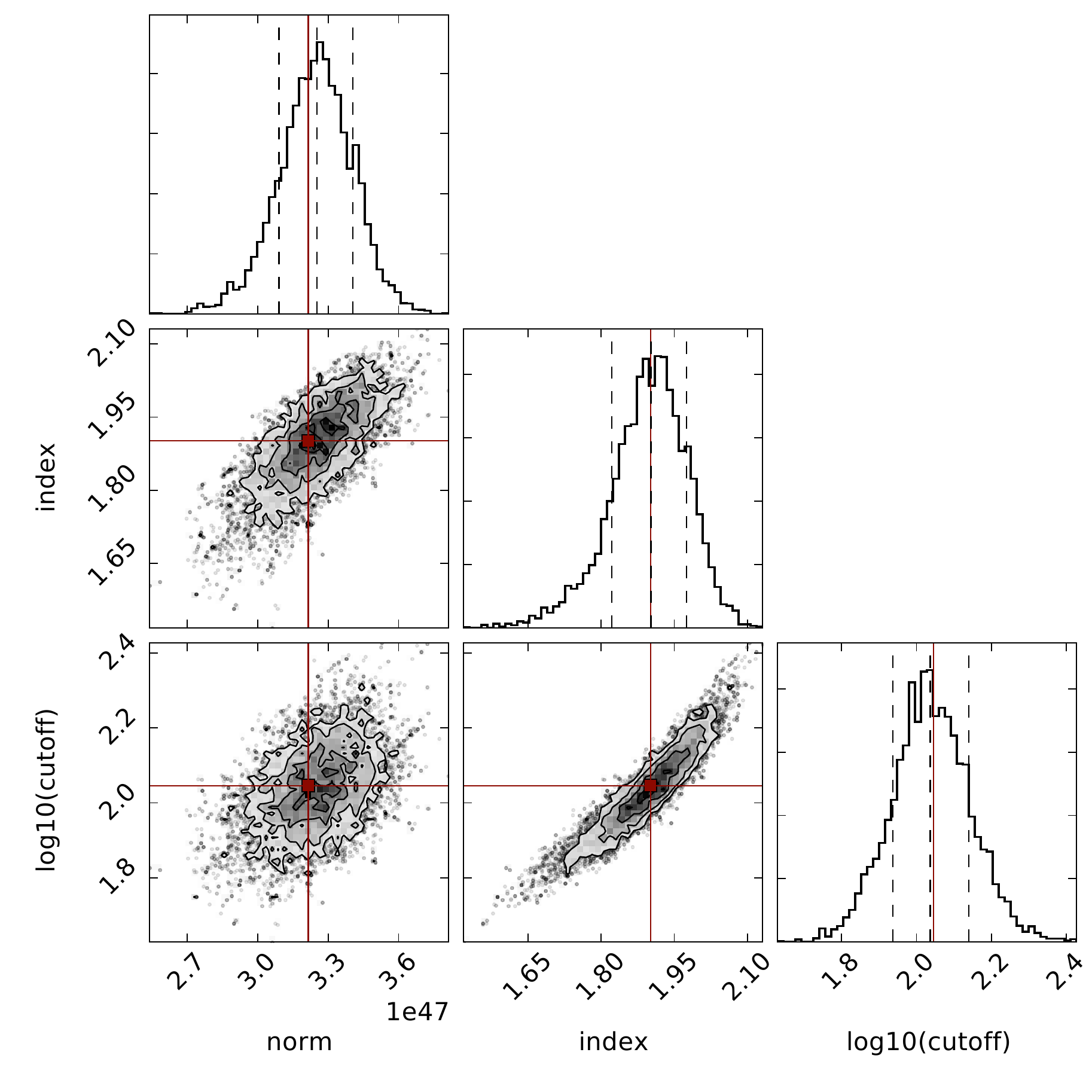}
    
    \end{center}
    \vspace{-2em}
    \caption{
        Distribution of the free model parameters.  \texttt{norm} is the
        particle distribution normalization at 5\,TeV  in units of
        $\mathrm{TeV}^{-1}$, \texttt{index} is the power law index, and
        \texttt{log10(cutoff)} is the decimal logarithm of the cutoff energy in
        units of TeV during the MCMC sampling run. The parameter vector with the
        highest likelihood found during the run is indicated by
        the red cross.
    }
    \label{fig:rxcorner}
\end{figure}

\begin{figure}[t]
    \begin{center}
        \includegraphics[width=0.9\textwidth]{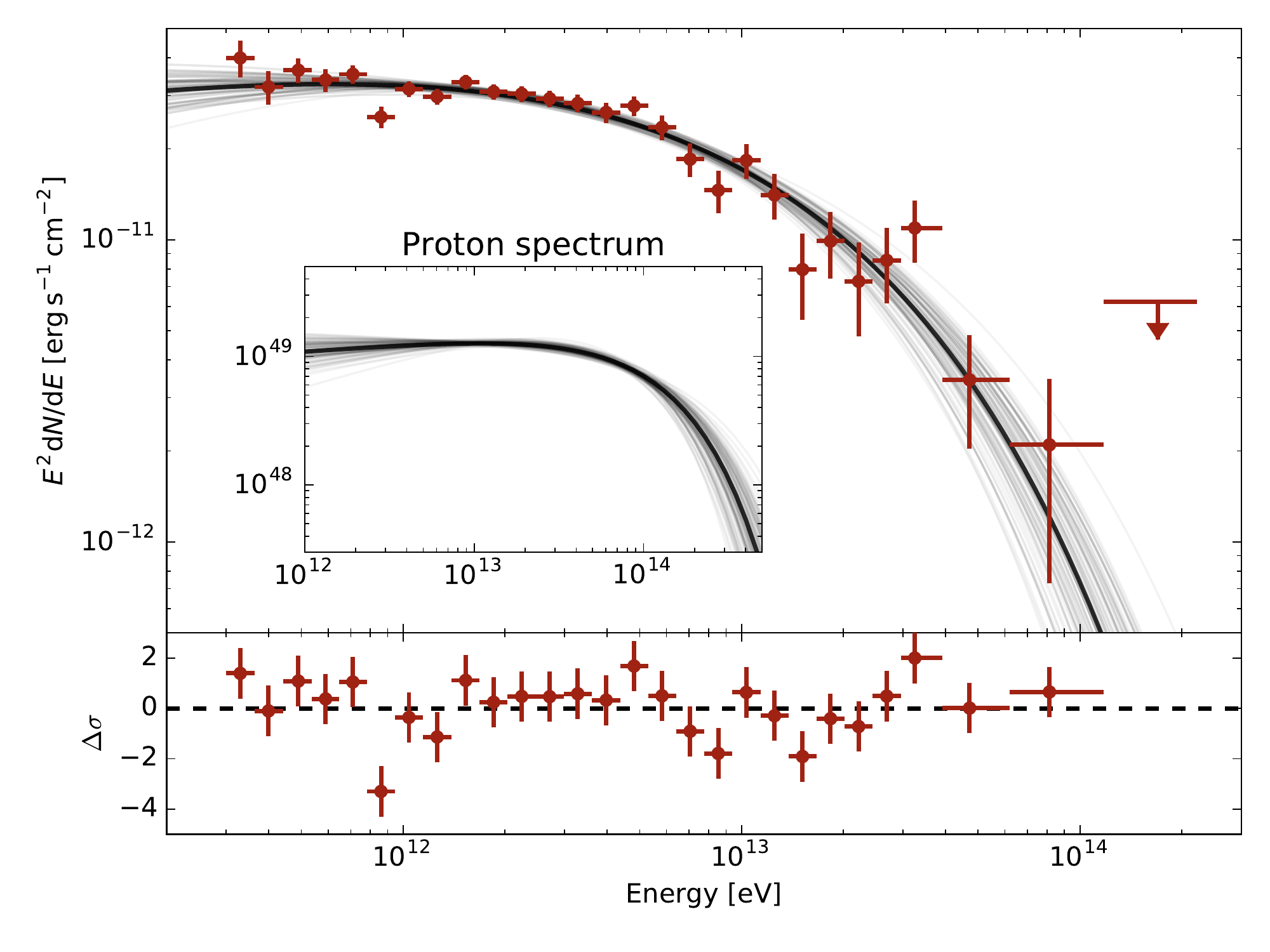}\\
    \end{center}
    \vspace{-2em}
    \caption{\hess\ spectrum of \rx\ \cite{2007A&A...464..235A}, computed spectrum
    from a hadronic model, and residuals of the maximum likelihood model (bottom
    panel). The thick black line indicates the maximum likelihood spectrum, and
    the gray lines are 100 samplings of the posterior distribution of the model
parameter vector. The inset shows the energy distribution of the proton
population in erg versus the proton energy in TeV.} \label{fig:rxmodel}
\end{figure}

The first step is the definition of the radiative model to be fit, and in this
case we will use the \texttt{ExponentialCutoffPowerLaw} particle distribution
function and \texttt{PionDecay} radiative model. The spectrum from
\cite{2007A&A...464..235A} can be read as a table with
\texttt{astropy.io.ascii}, and passed onto the \naima\ MCMC sampling functions
along with the defined model function. We run the sampling with 32 simultaneous
walkers (or sampling chains), for 100 steps of \emph{burn-in}, and 400 steps of
run that are saved for later analysis. The resulting sampled chains can be
analysed with a set of \naima\ functions that plot diagnostic and results plots.
Such a chain diagnostic produced with \texttt{naima.plot\_chain} can be seen for
the energy of the exponential cutoff in the particle distribution model in
Figure~\ref{fig:chain}.  Figure~\ref{fig:rxcorner} shows a \emph{corner} plot,
which plots the distribution for all parameters against each other. It can be
seen that there is a large positive covariance between the distributions of the
particle index and cutoff parameters. Finally, Figure~\ref{fig:rxmodel} shows a
comparison between the observed spectrum and the computed model, including the
spread of the spectrum derived from 100 random parameter vectors taken from the
MCMC chain, and an inset with the inferred particle energy distribution.

The result of the MCMC run provides estimates of the parameters of the
particle distribution function. For this spectrum and radiative model, the particle index is
constrained to $s=1.90\pm0.08$, and the cutoff energy is
$E_\mathrm{c}=110^{+30}_{-20}$\,TeV. The MCMC parameter chain can also be used to
compute the distribution of quantities derived from the particle spectrum, such
as total energy content in protons, which for \rx\ is constrained to
$W_p(E_p>1\,\mathrm{TeV}) = (5.74^{+0.17}_{-0.2})\times10^{49} n_\mathrm{H}^{-1}
d^{-2}_{1\,\mathrm{kpc}}$\,erg, where $n_\mathrm{H}$ is the target density and
$d_{1\,\mathrm{kpc}}$ is the distance to the source in kpc.

\section{Limitations and future development}\label{sec:future}

The main limitation of the approach used by \naima\ for spectral fitting is the
assumption of uncorrelated, gaussian errors. Even though this may be incorrect
for many spectra, mostly when considering fine structure, it is often the only
approach possible when simultaneously fitting published spectra from radio to
VHE gamma-rays. When instrument response functions are available, a way to avoid
this assumption is to use the \textsf{sherpa} models in
\texttt{naima.sherpa\_models}.

Future development of the package will focus on the
    addition of simple particle cooling functions (more complex physics,
        such as time-dependent particle evolution, should be done on a
        case-by-case basis), and use of \textsf{naima} radiative models in
        \textsf{gammapy} \cite{gammapy}, a Python
        package for gamma-ray data analysis.

\acknowledgments The author would like to thank F.~Aharonian, for providing the
motivation to develop a package such as \naima\ and make it public for general
use; D.~Khangulyan, E.~Kafexhiu, and G.~Vila for their discussions on radiative
model implementation; and C.~Deil for his contributions to the source code that
have helped make \naima\ more robust and user friendly. \naima\ makes use of
Astropy, a community-developed core Python package for Astronomy
\cite{2013A&A...558A..33A}, and matplotlib, a Python library for publication
quality graphics \cite{Hunter:2007}.

\providecommand{\href}[2]{#2}\begingroup\raggedright\endgroup


\begin{thebibliography}{10}

\bibitem{2001SPIE.4477...76F}
P.~{Freeman}, S.~{Doe}, and A.~{Siemiginowska}, {\it {Sherpa: a
  mission-independent data analysis application}},  in {\em Astronomical Data
  Analysis} (J.-L. {Starck} and F.~D. {Murtagh}, eds.), vol.~4477 of {\em
  Society of Photo-Optical Instrumentation Engineers (SPIE) Conference Series},
  pp.~76--87, Nov., 2001.
\newblock \href{http://arxiv.org/abs/astro-ph/0108426}{{\tt astro-ph/0108426}}.

\bibitem{2013A&A...558A..33A}
{Astropy Collaboration} et~al., {\it {Astropy: A community Python package for
  astronomy}},  {\em \aap} {\bf 558} (Oct., 2013) A33,
  [\href{http://arxiv.org/abs/1307.6212}{{\tt arXiv:1307.6212}}].

\bibitem{1970RvMP...42..237B}
G.~R. {Blumenthal} and R.~J. {Gould}, {\it {Bremsstrahlung, Synchrotron
  Radiation, and Compton Scattering of High-Energy Electrons Traversing Dilute
  Gases}},  {\em Reviews of Modern Physics} {\bf 42} (1970) 237--271.

\bibitem{2010PhRvD..82d3002A}
F.~A. {Aharonian}, S.~R. {Kelner}, and A.~Y. {Prosekin}, {\it {Angular,
  spectral, and time distributions of highest energy protons and associated
  secondary gamma rays and neutrinos propagating through extragalactic magnetic
  and radiation fields}},  {\em \prd} {\bf 82} (Aug., 2010) 043002,
  [\href{http://arxiv.org/abs/1006.1045}{{\tt arXiv:1006.1045}}].

\bibitem{2014ApJ...783..100K}
D.~{Khangulyan}, F.~A. {Aharonian}, and S.~R. {Kelner}, {\it {Simple Analytical
  Approximations for Treatment of Inverse Compton Scattering of Relativistic
  Electrons in the Blackbody Radiation Field}},  {\em \apj} {\bf 783} (Mar.,
  2014) 100, [\href{http://arxiv.org/abs/1310.7971}{{\tt arXiv:1310.7971}}].

\bibitem{1999ApJ...513..311B}
M.~G. {Baring}, D.~C. {Ellison}, S.~P. {Reynolds}, I.~A. {Grenier}, and
  P.~{Goret}, {\it {Radio to Gamma-Ray Emission from Shell-Type Supernova
  Remnants: Predictions from Nonlinear Shock Acceleration Models}},  {\em \apj}
  {\bf 513} (Mar., 1999) 311--338,
  [\href{http://arxiv.org/abs/astro-ph/9810158}{{\tt astro-ph/9810158}}].

\bibitem{1975ZNatA..30.1099H}
E.~{Haug}, {\it {Bremsstrahlung and pair production in the field of free
  electrons}},  {\em Zeitschrift Naturforschung Teil A} {\bf 30} (Sept., 1975)
  1099--1113.

\bibitem{2006PhRvD..74c4018K}
S.~R. {Kelner}, F.~A. {Aharonian}, and V.~V. {Bugayov}, {\it {Energy spectra of
  gamma rays, electrons, and neutrinos produced at proton-proton interactions
  in the very high energy regime}},  {\em \prd} {\bf 74} (Aug., 2006) 034018,
  [\href{http://arxiv.org/abs/astro-ph/0606058}{{\tt astro-ph/0606058}}].

\bibitem{2006ApJ...647..692K}
T.~{Kamae}, N.~{Karlsson}, T.~{Mizuno}, T.~{Abe}, and T.~{Koi}, {\it
  {Parameterization of {$\gamma$}, e$^{+/-}$, and Neutrino Spectra Produced by
  p-p Interaction in Astronomical Environments}},  {\em \apj} {\bf 647} (Aug.,
  2006) 692--708, [\href{http://arxiv.org/abs/astro-ph/0605581}{{\tt
  astro-ph/0605581}}].

\bibitem{2014PhRvD..90l3014K}
E.~{Kafexhiu}, F.~{Aharonian}, A.~M. {Taylor}, and G.~S. {Vila}, {\it
  {Parametrization of gamma-ray production cross sections for p p interactions
  in a broad proton energy range from the kinematic threshold to PeV
  energies}},  {\em \prd} {\bf 90} (Dec., 2014) 123014,
  [\href{http://arxiv.org/abs/1406.7369}{{\tt arXiv:1406.7369}}].

\bibitem{2003itil.book.....M}
D.~J.~C. {Mackay}, {\em {Information Theory, Inference and Learning
  Algorithms}}.
\newblock Oct., 2003.

\bibitem{2013PASP..125..306F}
D.~{Foreman-Mackey}, D.~W. {Hogg}, D.~{Lang}, and J.~{Goodman}, {\it {emcee:
  The MCMC Hammer}},  {\em \pasp} {\bf 125} (Mar., 2013) 306--312,
  [\href{http://arxiv.org/abs/1202.3665}{{\tt arXiv:1202.3665}}].

\bibitem{2007A&A...464..235A}
F.~{Aharonian} et~al., {\it {Primary particle acceleration above 100 TeV in the
  shell-type supernova remnant RX J1713.7-3946 with deep HESS observations}},
  {\em \aap} {\bf 464} (Mar., 2007) 235--243,
  [\href{http://arxiv.org/abs/astro-ph/0611813}{{\tt astro-ph/0611813}}].

\bibitem{2014MNRAS.445L..70G}
S.~{Gabici} and F.~A. {Aharonian}, {\it {Hadronic gamma-rays from RX
  J1713.7-3946?}},  {\em \mnras} {\bf 445} (Nov., 2014) L70--L73,
  [\href{http://arxiv.org/abs/1406.2322}{{\tt arXiv:1406.2322}}].

\bibitem{gammapy}
A.~Donath, C.~Deil, et~al., {\it Gammapy -- a python package for $\gamma$-ray
  astronomy},  {\em in press, Proceedings of the 34th ICRC} (2015).

\bibitem{Hunter:2007}
J.~D. Hunter, {\it Matplotlib: A 2d graphics environment},  {\em Computing In
  Science \& Engineering} {\bf 9} (2007), no.~3 90--95.

\end{thebibliography}
\end{document}